\documentclass[aps,prl,amsmath,amssymb,superscriptaddress,reprint,twocolumn]{revtex4}
\usepackage{epsfig,graphicx,color,calc,cancel}

\newcommand{\beq}{\begin{equation}}
\newcommand{\eeq}{\end{equation}}
\newcommand{\ba}{\begin{align}}
\newcommand{\ea}{\end{align}}
\renewcommand{\phi}{\varphi}

\usepackage{mathrsfs}   

 \let\o\omega

\begin{document}

\title{Finite-size scaling of the magnetization probability density \\ for the critical Ising model in slab geometry.}

\author{David Lopes Cardozo}
\author{Peter C. W. Holdsworth}
\affiliation{Laboratoire de Physique, \'Ecole Normale Sup\'erieure de Lyon, UMR CNRS 5672,
46 all\'ee d'Italie, 69007 Lyon, France}

\begin{abstract}
The magnetization probability density in d=2 and 3 dimensional Ising models in slab geometry of volume $L_{\parallel}^{d-1} \times L_{\perp}$ is computed through Monte-Carlo simulation at the critical temperature and zero magnetic field. The finite-size scaling of this distribution and its dependence on the system aspect-ratio $\rho=\frac{L_{\perp}}{L_{\parallel}}$ and boundary conditions is discussed. In the limiting case $\rho \to 0$ of a macroscopically large slab ($L_{\parallel} \gg L_{\perp}$) the distribution is found to scale as a Gaussian function for all tested system sizes and boundary conditions.
\end{abstract}

\maketitle

\section{Introduction}
\label{sec:intro}

\begin{figure}[b]
\center
\includegraphics[width=0.7\linewidth]{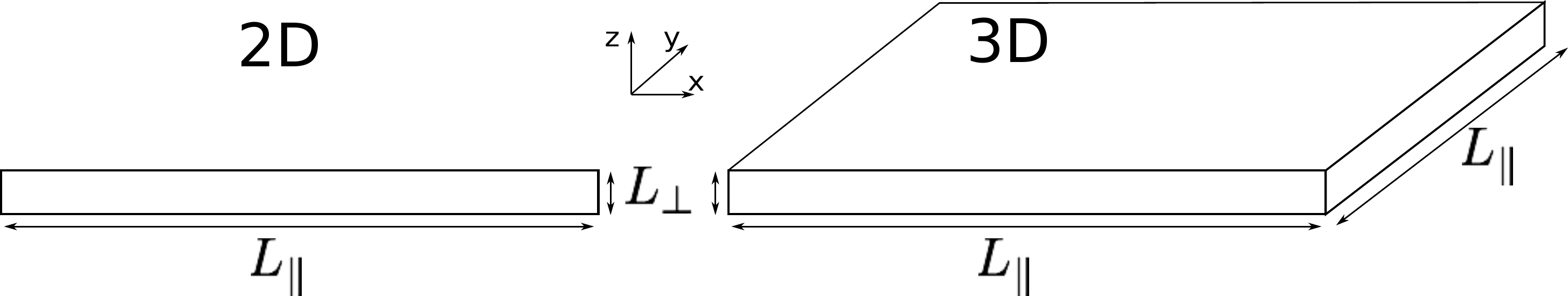}
\caption{The slab geometry in 2 and 3 dimensions.}
\label{slab_geom_2D_3D}
\end{figure}

Finite-size scaling (FSS) has proved to be a powerful tool for the interpretation and exploitation of the thermodynamics of confined critical systems \cite{binder_overcoming_2003}. The number one tool for computer modeling, FSS is finding a growing number of experimental applications as confinement on the nano and micro scales becomes more accessible. Examples include the critical Casimir effect  \cite{gambassi_casimir_2009,lopes_cardozo_critical_2014}, fluctuation spectra in confined liquid crystals \cite{joubaud_experimental_2008} or finite size order parameter scaling for superfluid helium films \cite{bramwell_phase_2015}. 
Here, we are interested in the form of fluctuations of the order parameter in confined critical two and three dimensional Ising models through the scaling behavior of the magnetization probability density. 
We make the transition between cubic and slab geometries, studying the influence of the aspect ratio on the fluctuations of the order parameter. 
This is motivated by the importance of the slab geometry for both theoretical and experimental applications of finite-size scaling. It is convenient in a system of dimensions $L_{\perp} \times L_{\parallel}^{d-1}$ (Fig.\ref{slab_geom_2D_3D}) to have a clearly identified confining length scale $L_{\perp}$ and $d-1$ non-confining ones such that  $L_{\parallel} \geq L_{\perp}$, this geometry being relevant for magnetic thin films experiments \cite{vaz_magnetism_2008} and wetting phenomenon \cite{binder_monte_2003}.

We start by presenting the case of fully periodic boundary conditions (BC) but, as one might expect intuitively,  the presence of boundary fields deeply influence the fluctuations. We study this influence in the following sections by using fixed $(+,\pm)$ boundary conditions in the confining direction $z$ and see that in the slab limit $L_\parallel \gg L_\perp$ the probability density of the magnetization tends to a Gaussian form whatever the boundary conditions. For a magnetic system, having $(+,\pm)$ boundaries amounts to applying a magnetic field, acting only on boundary spins, with the limiting  case of infinitely strong boundary fields corresponding to fixing the value of boundary spins to be either in the positive $(+)$ or negative $(-)$ directions. Thanks to the universality of finite size scaling \cite{pelissetto_critical_2002}, results obtained in a critical Ising model can be extended to other systems in the Ising universality class, including the liquid-gas phase transition and the demixing transition of binary fluids. In particular, binary fluids can be confined in wetting layers \cite{fukuto_critical_2005}, for which $(+,\pm)$ BCs correspond to boundaries which preferentially adsorb one of the two components.
%

\section[Form of the magnetization probability density]{Form of the magnetization probability density and finite-size scaling}
\label{sec:Form of the magnetization probability density}

At zero magnetic field and far from the critical temperature $T_c$, that it to say when the correlation length $\xi$ is small with respect to all confining length scales, two regimes can be identified \cite{amit_field_2005,binder_overcoming_2003}. At a temperature $T$ much higher than the critical temperature, thermal fluctuations dominate and the central limit theorem applies to the uncorrelated spins leading to a Gaussian distribution 
\beq
P(m,T,N) = \sqrt{\frac{N}{2\pi k_BT\chi(T)}}e^{-Nm^2/(2k_BT\chi)} \label{Pm1}
\eeq
for the probability density of the magnetization per spin $m$, with the magnetic susceptibility $\chi$ being independent of the total number of spins N \cite{landau_guide_2000}. At temperatures much lower than $T_c$, $P(m,T,N)$ is double peaked, symmetrically around 0, the positions of the peaks are N independent but their width evolves as $N^{-1/2}$ \cite{amit_field_2005}.
We are interested in the behavior close to $T_c$, in between these two limit cases. 

Equation (\ref{Pm1}) is a consequence of
 the very general statement that any normalized distribution  $\Pi$ of a stochastic quantity $x$ can be written in the scaling form, $\tilde{\Pi}(x/\sigma_x)$, such that
\begin{equation}
 \Pi(x) = (1/\sigma_x) \tilde{\Pi}(x/\sigma_x) ,
 \label{scaling0}
 \end{equation}
with $\sigma_x=\sqrt{\langle x^2 \rangle - \langle x \rangle^2}$ the standard deviation and  where $\langle . \rangle$ is a statistical average. For example, this statement is obviously verified if the distribution is a door function for which the width is $\sqrt{12} \sigma_x$ and amplitude $1/(\sqrt{12}\sigma_x)$ or a Gaussian $\Pi(x)= \frac{1}{\sqrt{2\pi \sigma_x^2}} \exp\left(-\frac{x^2}{2\sigma_x^2} \right)$. In this case, $x=m$ and  $\sigma^2=\langle m^2 \rangle - \langle m \rangle^2 =k_BT\chi/N$

In a thermodynamically large system, the correlation length diverges when the temperature approaches $T_c$ as $\xi=\xi_0^{\pm}t^{-\nu}$ with $\xi_0^{\pm}$ non universal amplitudes and $t=\frac{T-T_c}{T_c}$ the reduced temperature. In a finite system of hyper cubic geometry, such that $L_{\perp}=L_{\parallel}=L$, this divergence is cut-off by the confining size  $L$ and the ratio $L/\xi$ becomes a relevant finite size scaling parameter.  It then follows from Eq.\ref{scaling0} that the magnetization probability density function verifies~\cite{clusel_quelques_2005,bramwell_magnetic_2001} :
\begin{equation}
 P(m,L,t)  = (1/\sigma) \bar{P}(m/\sigma,L/\xi) ,
 \label{scaling2}
 \end{equation}
 %
in the scaling limit, where  the temperature dependence is now contained in $L/\xi$. The diverging susceptibility, $\chi=\chi_0|t|^{-\gamma}$ can be related to anomalous scaling for the width of the distribution, $\sigma^2 \propto L^{-(d-\gamma/\nu)}$ which, using the hyper scaling relation between critical exponents, $\gamma+2\beta=d\nu$, can be written as \cite{binder_finite_1981,clusel_quelques_2005,portelli_fluctuations_2002,portelli_universal_2001}:
\begin{equation}
\sigma \propto L^{-\beta/\nu},
\label{sigmaCubic}
\end{equation}
%
%
giving finally, the usual finite size scaling form for the distribution
\cite{binder_finite_1981,eisenriegler_helmholtz_1987,bramwell_magnetic_2001}
\begin{equation}
P(m,L,t)  = L^{\beta/\nu}\tilde{P}(mL^{\beta/\nu},tL^{1/\nu}).
 \label{scaling1}
\end{equation}
Critical exponents $\gamma$, $\beta$ and $\nu$ take their usual meaning throughout. The scaled distribution function $\tilde{P}$ is no longer necessarily Gaussian; it will depend both on dimension through the scaling exponent, $\beta$ and $\nu$ (for the 2D case $\beta=1/8$ and $\nu=1$ and for the 3D we take values $\beta=0.3265(3)$ and $\nu=0.6301(4)$ \cite{pelissetto_critical_2002}) and on boundary conditions. 

Much richer behavior occurs in systems with anisotropic confining geometry. We are interested in systems in slab geometry of dimensions $L_{\perp} \times L_{\parallel}^{d-1}$.
To clearly identify $L_{\perp}$ as the confining size, that is to say the one that will compete with $\xi$ and cut-off its divergence at the critical temperature, we shall keep in all cases $L_{\parallel} \geq L_{\perp}$. The aspect ratio $\rho=L_{\perp}/L_{\parallel}$ will govern the functional form of the distribution \cite{kaneda_shape_1999}, leading to a straight forward generalization of Eq.\ref{scaling2}:
\begin{equation}
P(m,L_{\perp},L_{\parallel},t)  = (1/\sigma) \hat{P}(m/\sigma,L_{\perp}/\xi, \rho) ,
\label{scaling3}
\end{equation}
with $\hat{P}(m/\sigma,\xi/L,1)=\tilde{P}(m/\sigma,L/\xi)$. 
In the limit ${L_{\parallel} \gg L_{\perp}}$, the argument that led to Eq.\ref{sigmaCubic} can be extended to the slab geometry : 
$\xi$ is proportionnal to $L_{\perp}$ in the scaling limit at criticality, so that the magnetic susceptibility scales as $\chi \propto L_{\perp}^{\gamma/\nu}$. Using the fluctuation-dissipation relation $\chi = N\sigma^2/k_BT$, with $N=L_{\perp}L_{\parallel}^{d-1}$ and the hyperscaling relation, we get:
\begin{equation}
\sigma = \hat{\sigma} L_{\perp}^{-\beta/\nu} \rho^{(d-1)/2}, 
\label{sigmad}
\end{equation}
which has been shown to hold in $d=2$ \cite{binder_finite-size_1989} and that we will show to work in $d=3$ in the following. 
The equations (\ref{scaling3}) and (\ref{sigmad}) lead us to propose the following scaling form for the magnetization distribution at $t=0$ when the limit $L_{\parallel} \gg L_{\perp}$ is approached :
\begin{widetext}
\begin{equation}
P(m,L_{\perp},L_{\parallel},t=0) = L_{\perp}^{\beta/\nu} \rho^{-(d-1)/2} \hat{P} \left(mL_{\perp}^{\beta/\nu} \rho^{-(d-1)/2},0,0 \right).
\label{scaling4}
\end{equation}
\end{widetext}
We predict the scaling function $\hat{P}(x,0,0)$ to be Gaussian
\begin{equation}
\hat{P}(x,0,0)=\frac{1}{\sqrt{2\pi \hat{\sigma}^2}} \exp\left(-\frac{x^2}{2\hat{\sigma}^2} \right) \ .
\label{Gaussian}
\end{equation}
The origin of the Gaussian form of the distribution at $T_c$ can be understood by stating that in the scaling limit with $L_{\parallel} \gg L_{\perp}$ the cut-off of the correlation length will be solely determined by the confining length $L_{\perp}$ so that the system can be divided in a number $N_\text{ind} \propto N/L_{\perp}^d$ of uncorrelated regions of volume $\xi^d \sim L_{\perp}^d$ \cite{portelli_universal_2001}.  In a sub-volume $L_{\perp}^d$ the magnetization fluctuates with $\sigma_{sv} \propto L_{\perp}^{-\beta/\nu}$ as stated in Eq.\ref{scaling1} \cite{binder_finite_1981}. The $N_\text{ind}$ sub-volumes being uncorrelated, the central-limit theorem tells us that the total magnetization will tend to have Gaussian fluctuations of standard deviation $\sigma=\sigma_{sn}/\sqrt{N_\text{ind}}$, as $N_{\text ind}$ is increased. 
However, the scaling standard deviation $\hat{\sigma}$ is a non-universal quantity, it relates to non-universal scaling amplitudes of $\chi$ and $\xi$, and depends on the boundary conditions, as we will see in the following.
%

\section{Numerical method}

We have tested numerically these predictions for the scaling form of the probability density through Monte-Carlo simulation. We recall the Ising model Hamiltonian:
\begin{equation}
\mathscr{H} = -J \sum\limits_{ <i,j> } S_i S_j,
\end{equation}
where the sum runs over all pairs of nearest neighbors, $J$ is the coupling constant set here to $J=1$ for convenience and $S_i$ are spins of value $\pm 1$. The spins are set on a hyper cubic lattice of dimensions $L_{\parallel}^{d-1} \times L_{\perp}$ and were updated through a hybrid Wolff/Metropolis algorithm \cite{newman_monte_1996,wolff_collective_1989} to reduce critical slowing down. By counting the number of occurrence $\mathscr{N}(m)$ of each possible value $m$ of the reduced magnetization 
\beq
\hat{m}= \frac{1}{N} \sum\limits_{i=1}^{N}  S_i \ , 
\eeq
where $N=L_{\perp}L_{\parallel} ^{d-1}$ is the total number of spins, we can estimate the probability for the system to exhibit a reduced magnetization $m$
\begin{equation}
P(m) = \frac{1}{Z} \sum\limits_{\left\{ S_i \right\} } e^{-\beta \mathscr{H}} \delta(m-\hat{m}),
\end{equation}
where $Z=\sum\limits_{\left\{ S_j \right\} } e^{-\beta \mathscr{H}}$ is the partition function, and the sum runs over all possible spin configurations. The total number of spins $N$ in the system fixes the discrete number of values the magnetization can take so that
\beq
P(m)=\frac{N}{2}\mathscr{N}(m) \ ,
\eeq
ensuring that the normalization $\int\limits_{-1}^{1}P(m) \mathrm{d}m=1$ is conserved whatever the system size. 

In two dimensions, only the case of fully periodic BC was investigated. In three dimension, periodic BC were always set in the $x,y$ directions (Fig.~\ref{slab_geom_2D_3D}) while in the  confining $z$ direction either periodic, or fixed BC were used. Fixed boundary conditions were realized by considering that the system is confined in the $z$ direction between two layers of spins with fixed values, either positive $(+)$ or negative $(-)$. We use the convention that $L_\perp$ refers to the number of layers of fluctuating spins, whatever the boundary conditions \cite{hasenbusch_thermodynamic_2010,hasenbusch_thermodynamic_2011,hasenbusch_thermodynamic_2015}.
Both the symmetric $(++)$ and anti-symmetric $(+-)$ cases were investigated.

\section{2D systems with fully periodic boundary conditions}

\begin{figure}[b]
\centering
{a) \hspace{-0.25cm}} 
\includegraphics[width=0.46\textwidth]{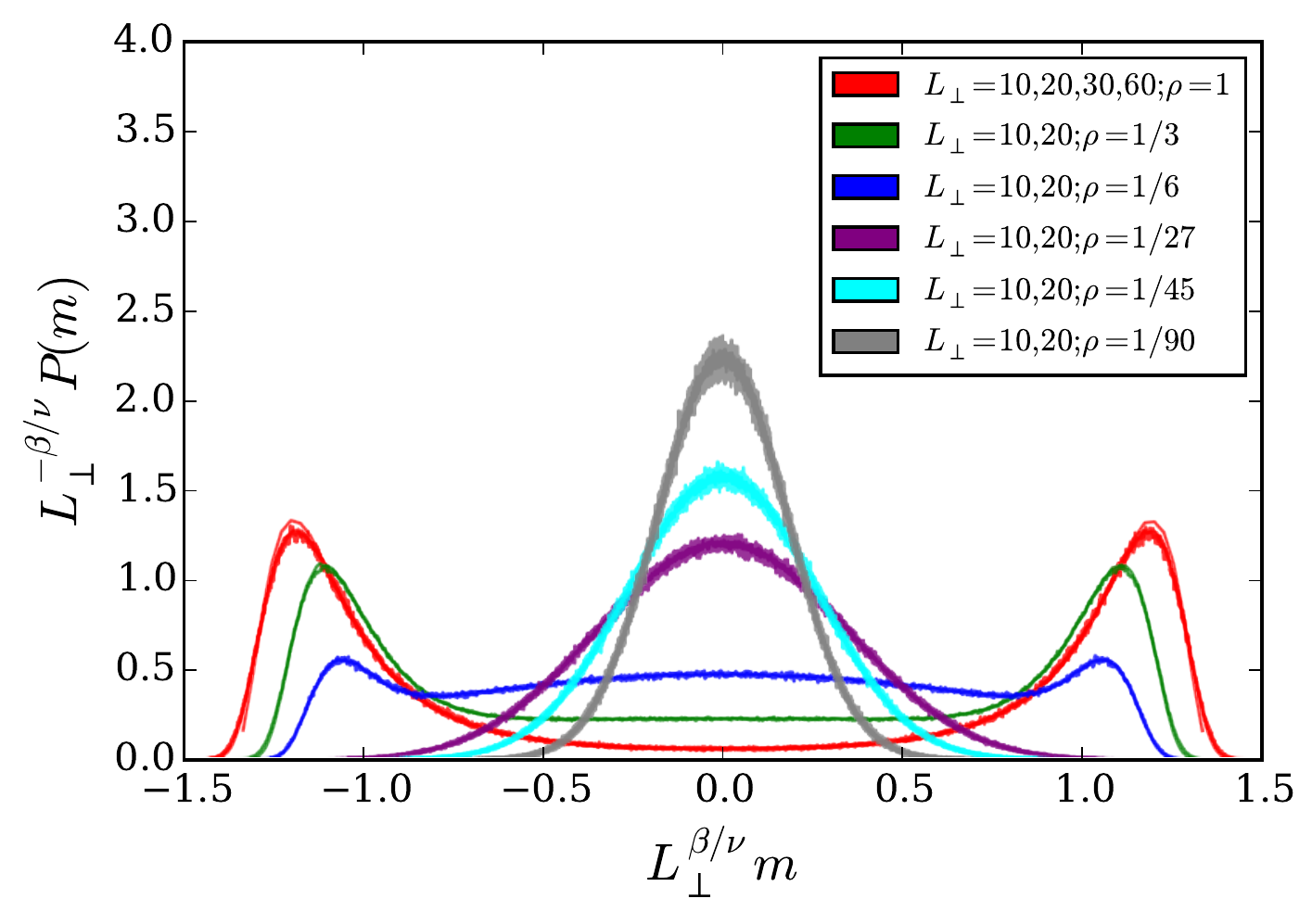}
{b) \hspace{-0.25cm}} 
\includegraphics[width=0.46\textwidth]{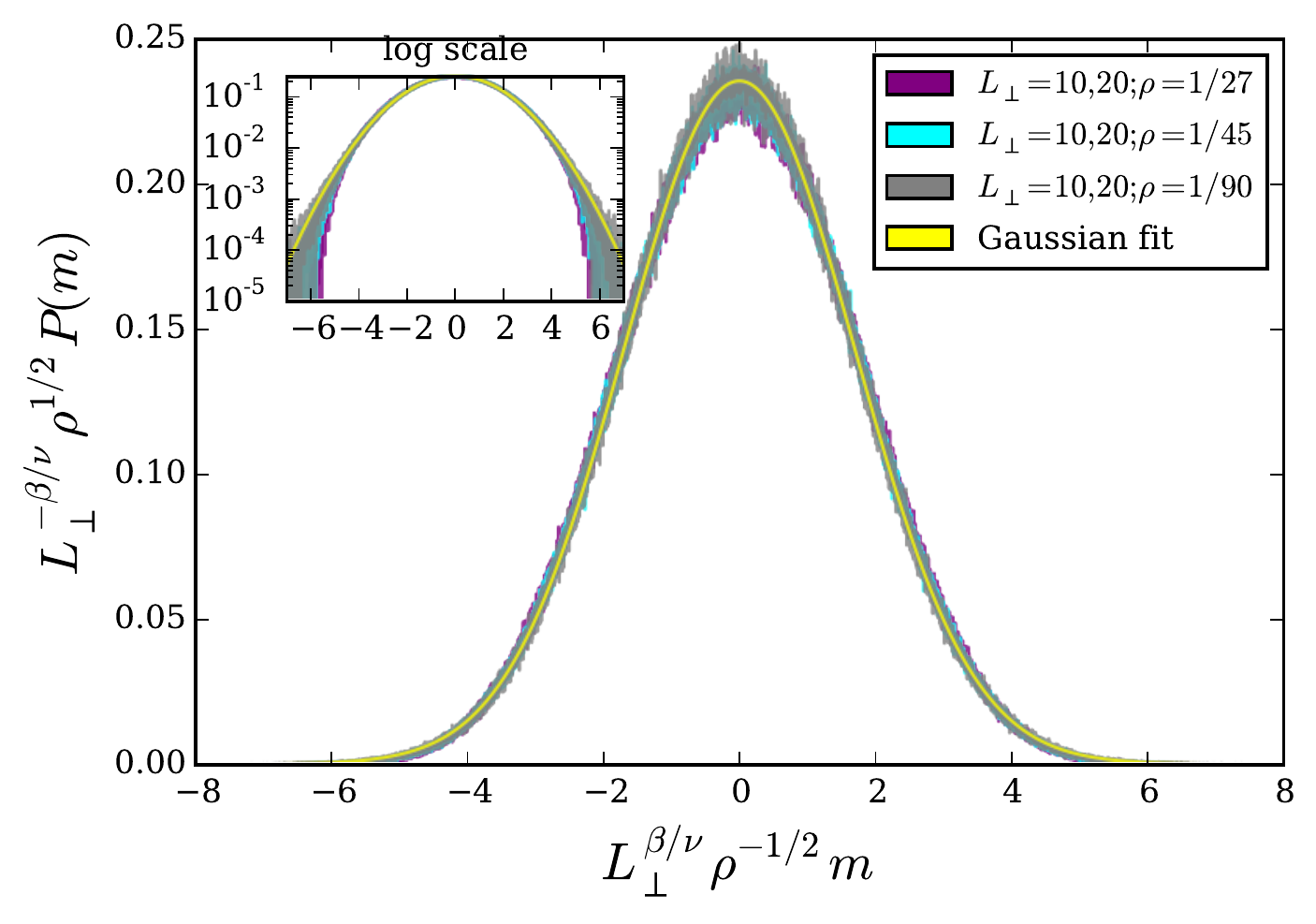}
\caption{ 
{\bf a)}  $L_\perp^{-\beta/\nu}P$ vs $mL_\perp^{\beta/\nu}$ with $P(m)$ obtained in a 2D Ising system with fully periodic boundary conditions at the critical temperature $T=T_c^{2D}$. The curves for systems with the same aspect ratio $\rho$ collapse \cite{kaneda_shape_1999}. 
As the aspect ratio is changed from the square case $\rho =1$ towards the limit $L_{\parallel} \gg L_{\perp}$, the functional form of the distribution goes from a bimodal to a unimodal one. 
{\bf b)} $L_{\perp}^{-\beta/\nu} \rho^{1/2}P$ vs $mL_{\perp}^{\beta/\nu} \rho^{-1/2}$ displayed only for unimodal distributions, validating the expected scaling from Eq.\ref{scaling4} in the limit  $L_{\parallel} \gg L_{\perp}$. The continuous line is a Gaussian fit which proves excellent \cite{binder_finite-size_1989}. {\bf Inset:} same, with logarithmic scale.}
\label{Pred_Tc_2D}
\end{figure}

We computed the probability density $P(m)$ at the critical temperature $T_c^{2D}$ for different confining dimensions $L_{\perp} \in [10:60]$ and aspect ratios, ranging from the square case $\rho=1$ to $\rho=1/100$. As can be seen in Figure \ref{Pred_Tc_2D}a) the aspect ratio affects the functional form of the distribution which goes from a bimodal one for square systems to a unimodal form as we get closer to the limit $\rho \to 0$. 
Plotting $L_\perp^{-\beta/\nu}P$ as a function of $mL_\perp^{\beta/\nu}$ gives a convincing collapse of distributions obtained for systems with similar aspect ratios, as expected from \cite{kaneda_shape_1999} and from Eq.\ref{scaling1} and Eq.\ref{scaling3}.  In the square case the collapse is imperfect at the maxima of the distribution for the smallest system sizes studied $L_\perp=10$ signaling corrections to the scaling limit.

%
\begin{figure}[h!]
\centering
\includegraphics[width=0.47\textwidth]{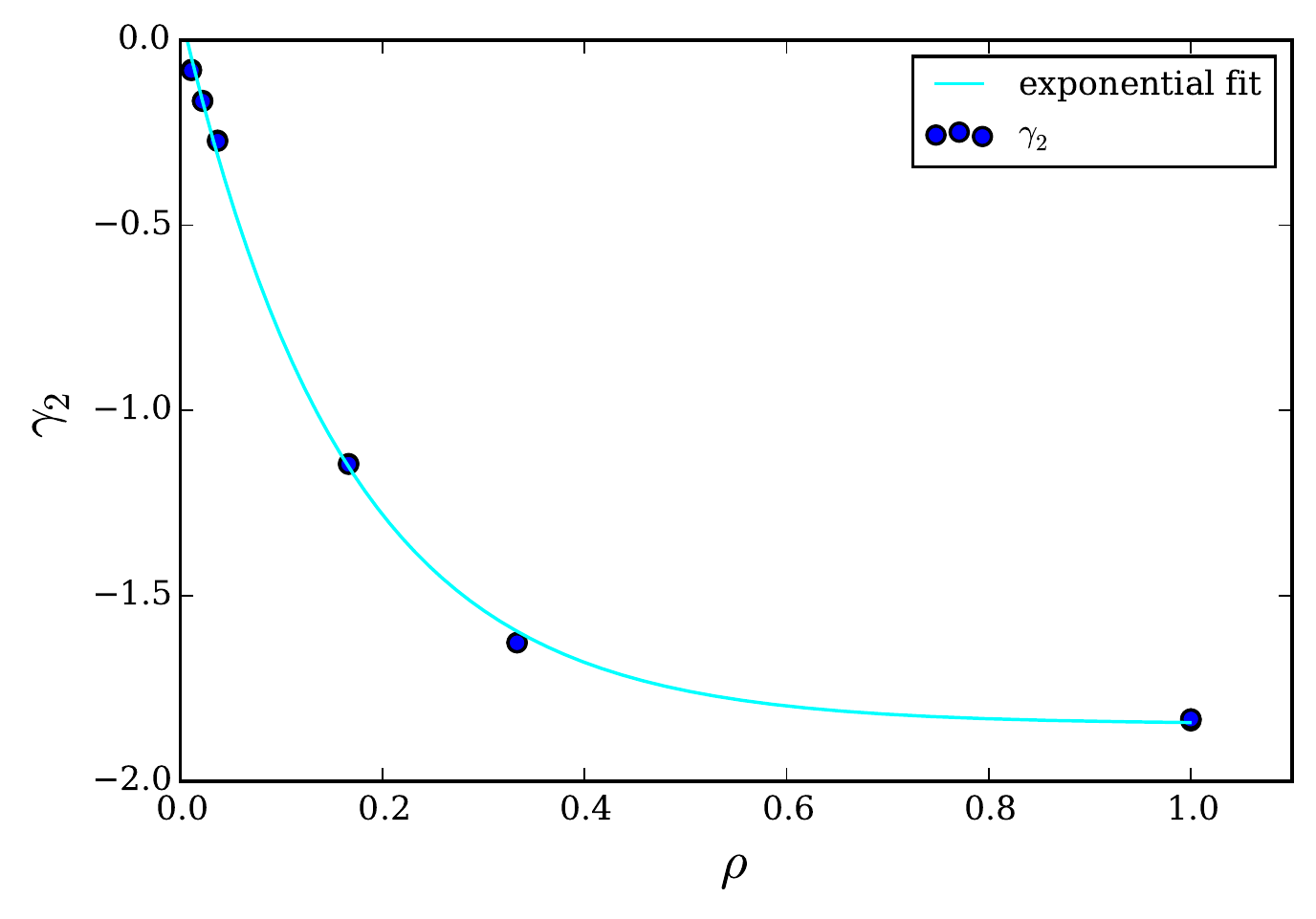}
\caption{ 
Kurtosis $\gamma_2$ of all the distributions of Fig.\ref{Pred_Tc_2D} as a function of $\rho$. The continuous line is an exponential fit intended as a guide to the eye. The kurtosis does not depend on the confinement $L_\perp$ but only on the aspect ratio $\rho$.
}
\label{Pred_Tc_2Dbis}
\end{figure}

%
Particularizing systems with aspect ratios $\rho \leq 1/27$ for which the distribution has reached a unimodal form, we can see in figure \ref{Pred_Tc_2D}b) that the expected Gaussian behavior in the limit $L_{\parallel} \gg L_{\perp}$ \cite{binder_finite-size_1989} and the scaling form of Eq.\ref{scaling4} are very well verified.
We stress that in this limit, all the magnetization distributions collapse according to Eq.\ref{scaling4}, not just those for systems with similar aspect ratios. 
A Gaussian fit (Eq.\ref{Gaussian}) performed on the reduced distribution for the biggest system size available proved excellent, with a standard deviation $\hat{\sigma}=1.6835(35)$. We recall that a Gaussian distribution is entirely characterized by its two first moments (the mean and the standard deviation). Here, with periodic boundary conditions, the third moment is zero by symmetry, so that the lowest order moment of the distributions characterizing a deviation from a Gaussian form is the fourth, or kurtosis:
\begin{equation}
\gamma_2=\frac{\langle (m- \langle m \rangle)^4\rangle}{ \sigma^2} -3  ,
\label{kurtosis}
\end{equation}
shown in figure~\ref{Pred_Tc_2Dbis}. The excess kurtosis is proportional to the Binder Cumulant $U_4=-3\gamma_2$  \cite{binder_finite_1981} which, in the scaling limit, is used to estimate the critical temperature as it is a universal constant with respect to the system size $L_\perp$ at $T_c$. This is coherent  with our observation that the kurtosis does not depend on the confinement $L_\perp$ within our current precision (Fig.\ref{Pred_Tc_2Dbis}). The kurtosis only depends on the aspect ratio $\rho$.
As $\rho$ is decreased the kurtosis goes to zero, indicating an evolution towards a Gaussian form of the distribution when reaching the limit of slab geometry, as we have predicted.
%
%

\section{3D systems with fully periodic boundary conditions}


Different BCs were used in the confining direction $\hat{z}$ but in order to extend the results of the previous section to the 3D case we start with the fully periodic boundaries. 

\begin{figure*}[!thbp]
\centering
{a) \hspace{-0.2cm}} 
\includegraphics[width=0.47\textwidth]{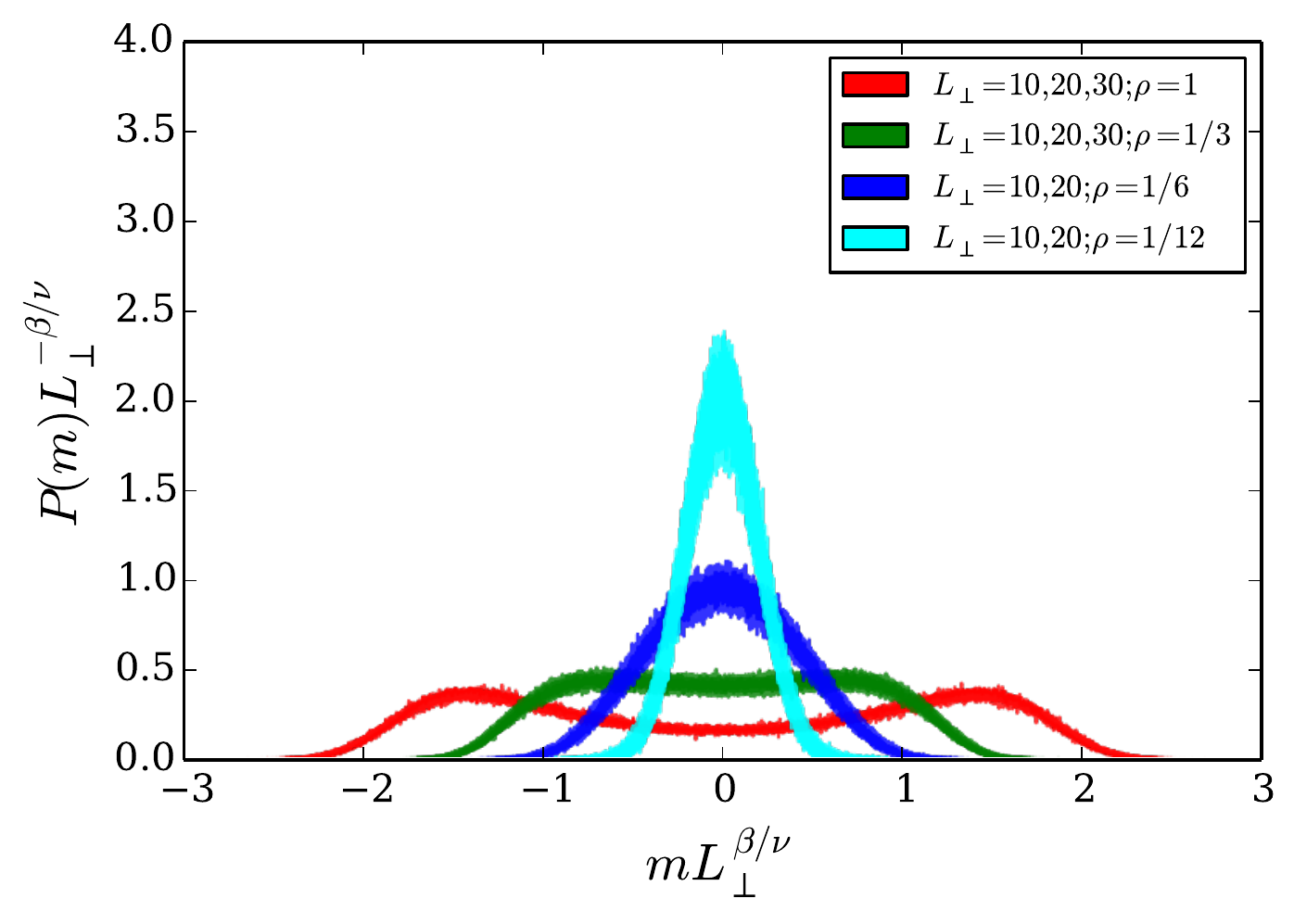}
{b) \hspace{-0.2cm}} 
\includegraphics[width=0.47\textwidth]{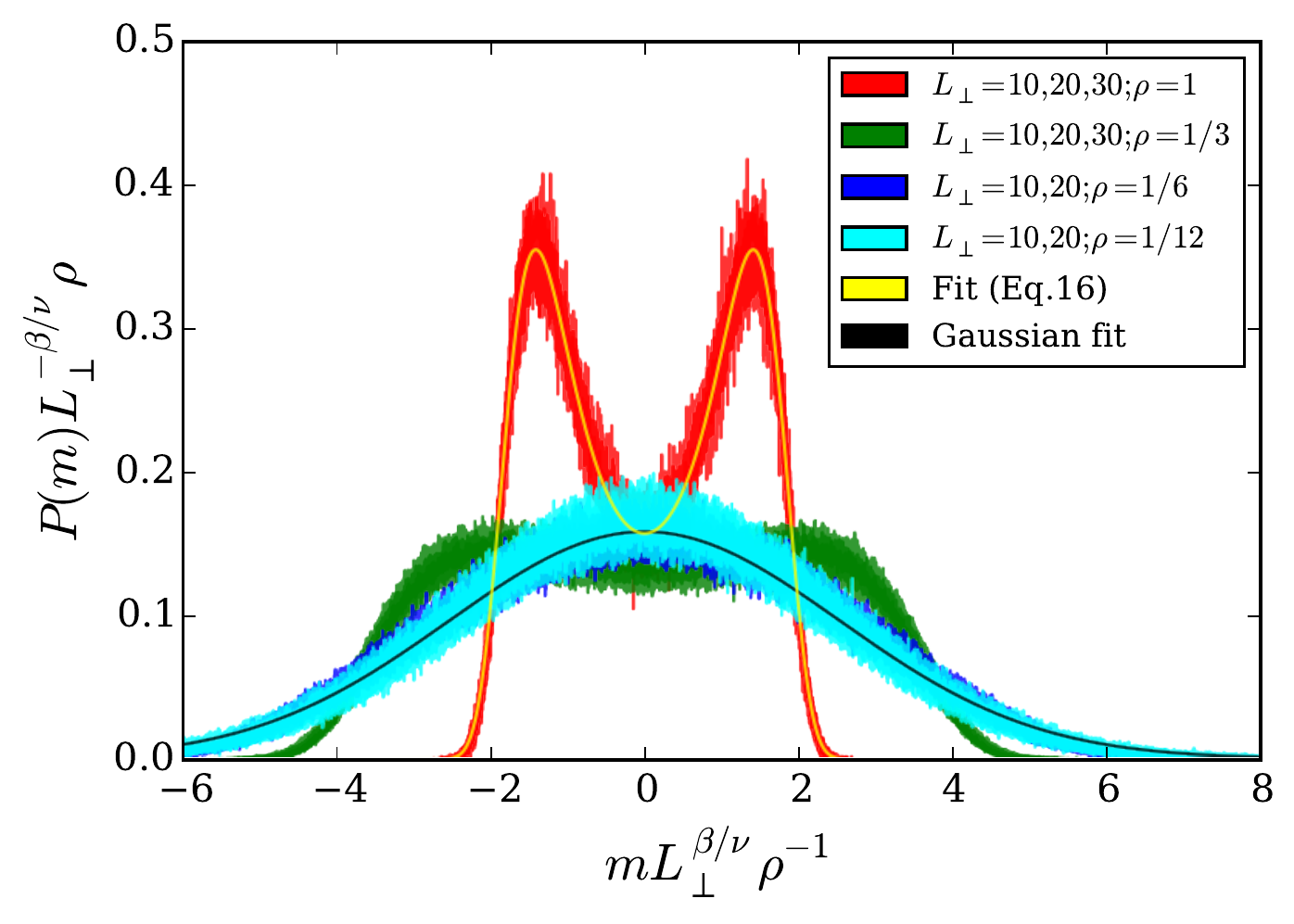}

{c) \hspace{-0.2cm}} 
\includegraphics[width=0.47\textwidth]{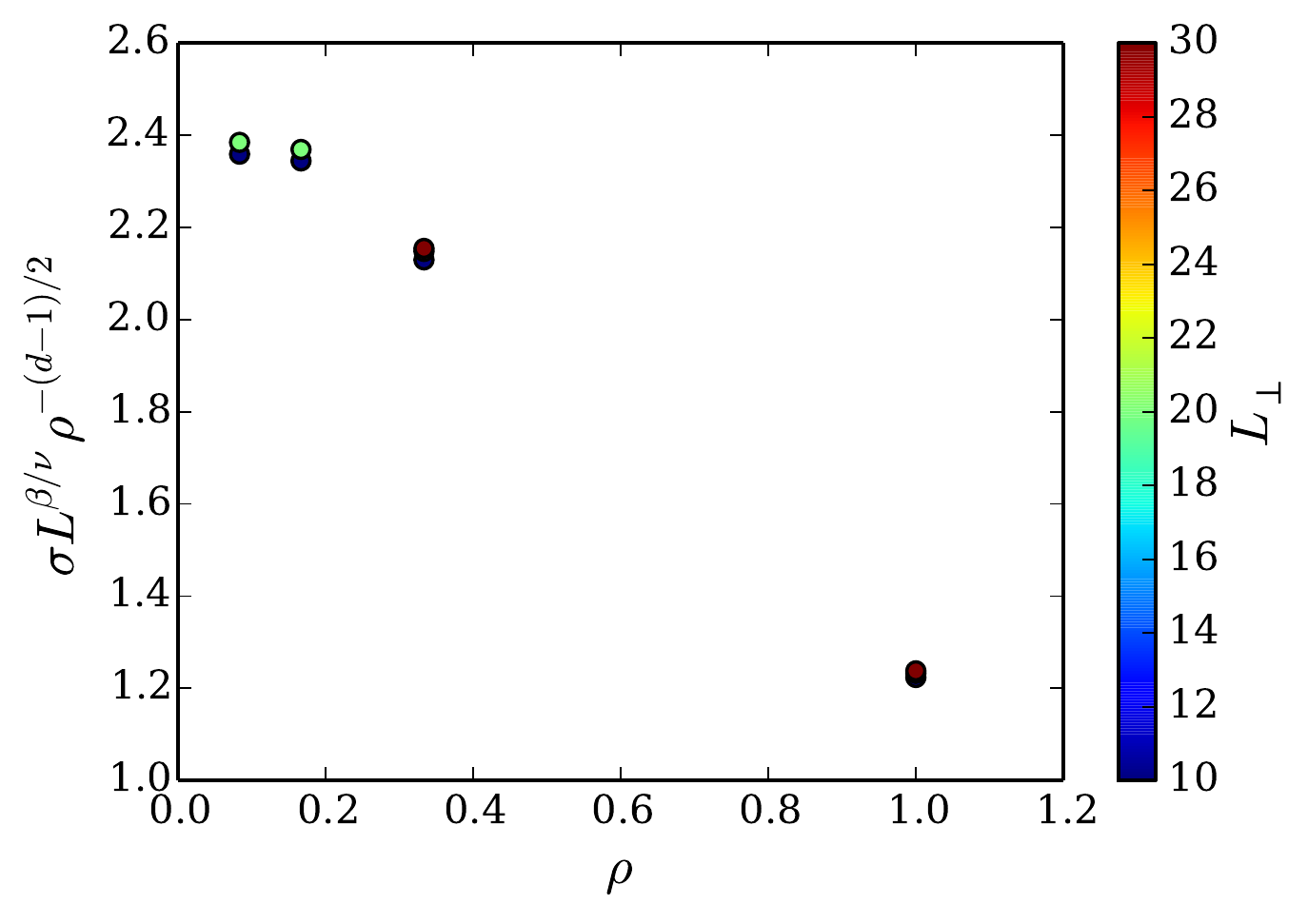}
{d) \hspace{-0.2cm}} 
\includegraphics[width=0.47\textwidth]{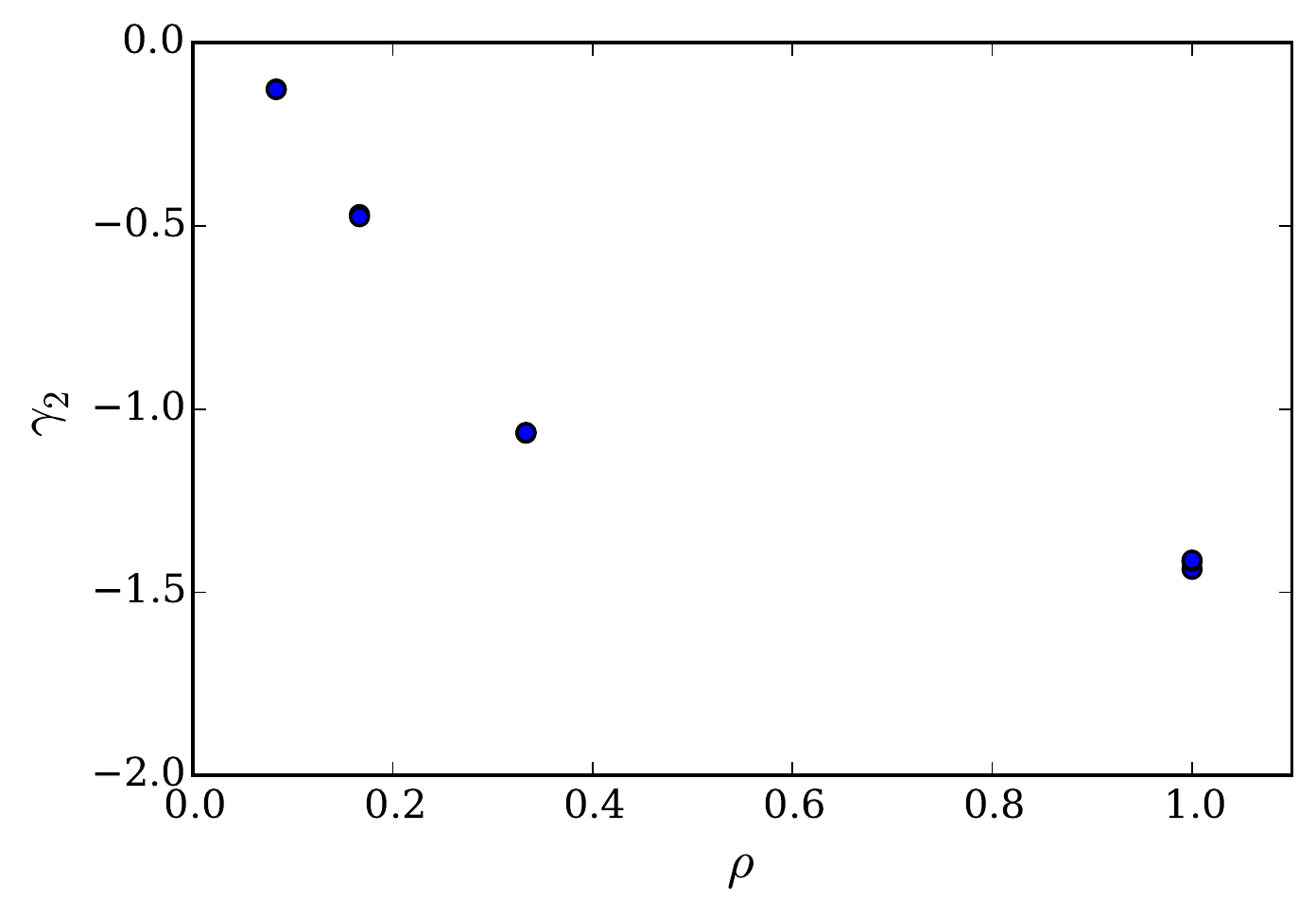}

\caption{Results obtained by Monte-Carlo simulation of 3D Ising system with fully periodic boundary conditions, for various system thicknesses $L_{\perp}$ and aspect ratios $\rho$ at the critical temperature $T=4.5115=T_c^{3D}$.
{\bf a)} $L_{\perp}^{-\beta/\nu}P $ as a function of $mL_{\perp}^{\beta/\nu}$. As the aspect ratio $\rho$ is changed from 1 to $1/12$ the distribution evolves from a bimodal to a unimodal functional form. Distributions for system with similar aspect ratios collapse, but the functional form of the collapse depends on the aspect ratio.
{\bf b)} $L_{\perp}^{-\beta/\nu}P$ as a function of $mL_{\perp}^{\beta/\nu}$. For distributions in the cubic case $\rho=1$, we obtain an excellent collapse and the yellow continuous line is a fit of an Ansatz proposed in ref.~\cite{tsypin_probability_2000} (see main text). Distribution for both $\rho=1/6$ and $1/12$ which reached a unimodal form collapse. The black continuous line is Gaussian fit.
{\bf c)} $L_{\perp}^{-\beta/\nu} \rho \sigma$ tends to a saturation value $\hat{\sigma}$ as the slab limit $\rho \to 0$ is approached.
{\bf d)} Kurtosis $\gamma_2$ of all the distributions as a function of $\rho$.
}
\label{P_3D_slab_PBC}
\end{figure*}

With fully periodic boundary conditions, the evolution of $P(m)$ with the aspect ratio at the critical temperature in three dimensions is qualitatively equivalent to the two dimensional case. As figure \ref{P_3D_slab_PBC}a) shows, when  $\rho$ evolves from 1 to $1/12$, the distribution evolves from a bimodal to a unimodal functional form \cite{kaneda_shape_1999}. Equivalently to the 2D case, Eq.\ref{scaling3} is verified:  plotting $L^{-\beta/\nu}P$ as a function of $mL^{\beta/\nu}$ gives a convincing collapse of distributions for systems with the same aspect ratio. The functional form of the collapse evolves with $\rho$ from bimodal for $\rho=1$ to Gaussian as $\rho\rightarrow 0$.

In the case of cubic $\rho=1$ systems, a high resolution study of the Ising and spin-1 Blume-Capel models \cite{tsypin_probability_2000} has shown the functional form:
\begin{equation}
\resizebox{1.\hsize}{!}{$P(m) \propto \exp\left(-\left(\frac{M^2}{M_0^2L^{-2\beta/\nu}}-1\right)^2\left(a\frac{M^2}{M_0^2L^{-2\beta/\nu}}+c\right)\right)$} \ ,
\label{PfunctionM}
\end{equation}
with $L=L_{\perp}=L_{\parallel}$ and a, c, $M_0$ non-universal factors which may contain corrections to scaling, to accurately fit the probability density. This result can be made consistent with the scaling hypothesis of Eqn.\ref{scaling1} by using the form
%
\begin{equation}
\resizebox{1.\hsize}{!}{$P = \frac{P_0}{L^{-\beta/\nu}}\exp\left(-\left(\frac{M^2}{M_0^2L^{-2\beta/\nu}}-1\right)^2\left(a\frac{M^2}{M_0^2L^{-2\beta/\nu}}+c\right)\right)$},
\end{equation}
%
in which we fitted $a,c,P_0,M_0$ to give excellent agreement with numerical data, see Fig.~\ref{P_3D_slab_PBC}b).

Figure \ref{P_3D_slab_PBC}b) shows $L_{\perp}^{-\beta/\nu}\rho^{(d-1)/2}P $ as a function of $mL_{\perp}^{\beta/\nu}\rho^{-(d-1)/2}$ following the scaling form proposed in Eq.\ref{scaling4}. Unimodal distributions, that is to say for $\rho \leq 1/6$, collapse whatever the aspect ratio. A Gaussian fit (Eq.\ref{Gaussian}) of the reduced distribution for the biggest system available indicates that the Gaussian behavior predicted in 2D in the limit $L_{\parallel} \gg L_{\perp}$ \cite{binder_finite-size_1989} indeed holds in 3D, with $\hat{\sigma} \approx 2.54$. 
Thus, the scaling form proposed in Eq.\ref{scaling4} appears verified in 3D as it was in 2D. Yet, the collapse is a little less convincing, probably because we could not as easily reach large values for $L_\parallel$. Nevertheless, the Gaussian behavior clearly is approached in the limit $\rho \to 0$. Following Eq.\ref{sigmad}, approaching this limit means having $L_{\perp}^{-\beta/\nu} \rho \sigma$ approaching a non universal constant $\hat{\sigma}$. This can be verified by looking at the dependency of $L_{\perp}^{-\beta/\nu} \rho \sigma$ on $\rho$, Fig.~\ref{P_3D_slab_PBC}c), which indeed displays a saturation close to $\rho=0$ at a value close to $\hat{\sigma} \approx 2.4$.
Moreover, as in 2D, we used the kurtosis to characterize the evolution of the distribution towards a Gaussian when reaching the limit of slab geometry, Fig.~\ref{P_3D_slab_PBC}d). The kurtosis depends  slightly on the confinement $L_\perp$ in the cubic $\rho=1$ case (corrections to scaling) but strongly on the aspect ratio, approaching zero as $\rho \to 0$, confirming that we approach Gaussian behavior.
%
%

\section{Influence of fixed boundary conditions}

\subsection{Fixed $(+-)$ boundary conditions}

\begin{figure*}[!t]
\vspace{-0.3cm}
\centering
{a) \hspace{-0.25cm}} 
\includegraphics[width=0.47\textwidth]{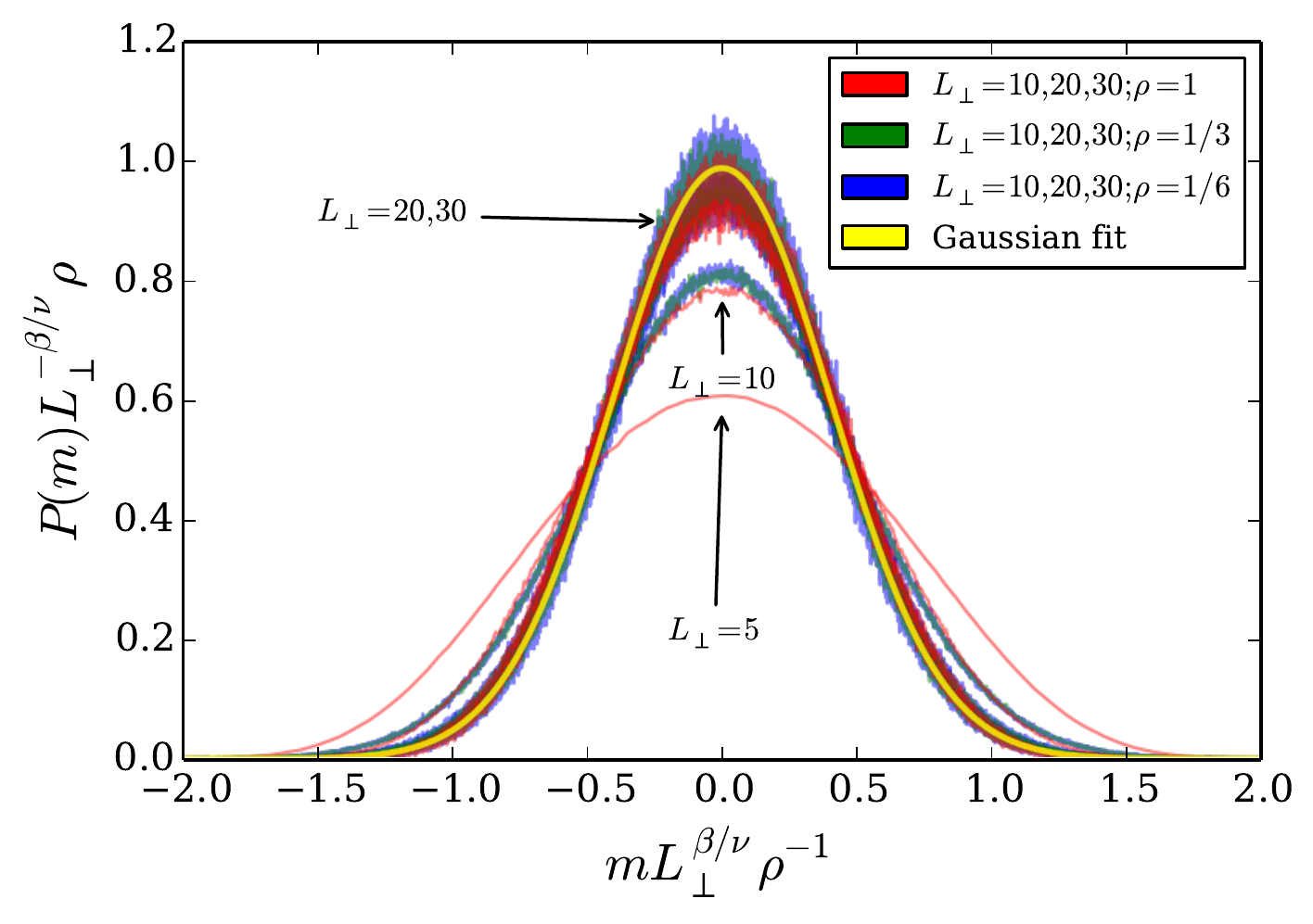}
\includegraphics[width=0.45\textwidth,height=6cm]{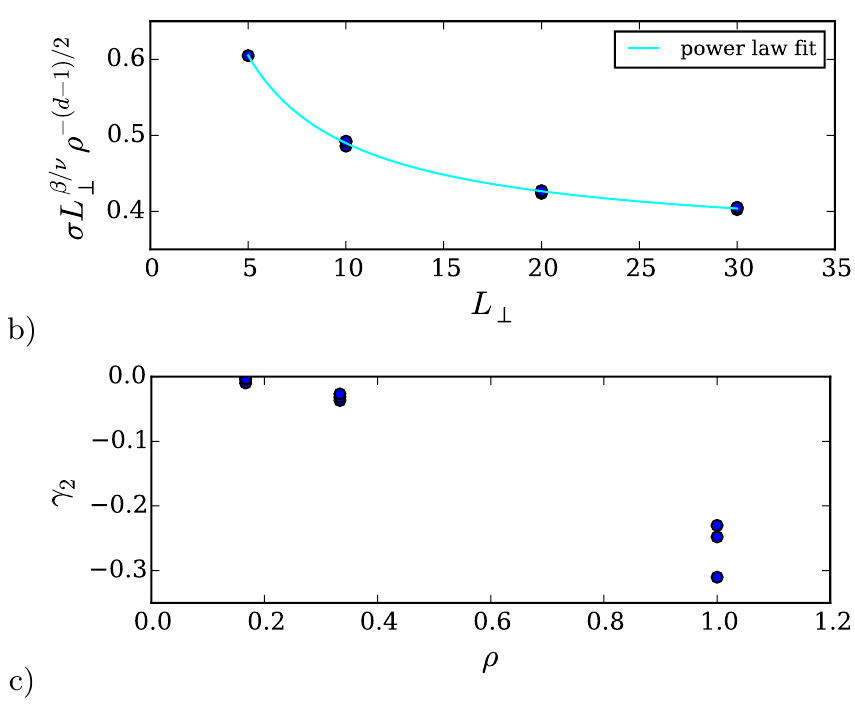}
\vspace{-0.1cm}
\caption{Results of Monte-Carlo simulation of 3D Ising systems with $(+-)$ BC for various thicknesses $L_{\perp}$ and aspect ratios $\rho$ at the critical temperature $T_c^{3D}$.
{\bf a)} $L_{\perp}^{-\beta/\nu} \rho P$ versus $mL_{\perp}^{\beta/\nu} \rho^{-1}$. All distributions are unimodal.  We obtain a very convincing collapse of all data for thicknesses $L_\perp \geq 20$, whatever the aspect ratio. The scaling functions obtained for thicknesses $L_{\perp}=5,10$ have greater variances. The continuous line is a  Gaussian fit of the distribution for the biggest available system.
{\bf b)} $L_{\perp}^{\beta/\nu} \rho^{-1} \sigma$ as a function of $L_\perp$. The continuous line is a power law fit of form $\hat{\sigma} + b_{eff} L_\perp^{-\omega_{eff}}$, intended as a guide to the eye.
{\bf c)} Kurtosis of all the distributions as a function of $\rho$. The kurtosis depends mainly on $\rho$ and goes to zero in the slab limit $\rho \to 0$. It also depends in a less pronounced way on the confinement $L_\perp$, especially when $\rho=1$. 
}
\label{P_3D_BCMP}
\end{figure*}

The presence of fixed boundary conditions deeply affects the behavior of the system. Fixed anti-symmetric $(+-)$ BC in the confining direction act as local magnetic fields and put a topological constraint on the magnetization that induces a boundary between a positively and a negatively magnetized region. This is similar to, but not quite the same, as a phase separation, induced by fixing the total magnetization. In that case, fluctuations are suppressed at large length scales by the constraint that  $m$ must remain a constant. Here, the magnetization is free to vary despite the topological domain wall imposed by the $(+-)$ boundaries, allowing global fluctuations.
Experimentally, the situation with $(+-)$ BC can be achieved for a critical binary polymer mixture of components A and B in a wetting layer or by confining it between two plates with each surface preferentially adsorbing one of the components of the mixture. The surfaces thus induce a phase boundary between an A rich phase and a B rich phase. We therefore expect that the average magnetization of systems with $(+-)$ BC will fluctuate around 0 as a result of fluctuations of the phase interface. This is indeed the case as can be seen in figure \ref{P_3D_BCMP}a): the magnetization probability density remains unimodal at the critical temperature whatever the aspect ratio $\rho$, ranging from 1 to $ 1/9$, and the system thickness $L_{\perp}$.

%
%
Plotting $L_{\perp}^{-\beta/\nu} \rho P$ as a function of $mL_{\perp}^{\beta/\nu} \rho^{-1}$ the scaling form of Eq.\ref{scaling4} leads to a convincing collapse for systems with $L_\perp \geq 20$ and all the studied aspect ratios, the master curve being well fitted by a Gaussian of  standard deviation $\hat{\sigma}=  0.43(4)$. The scaling functions obtained for thicknesses $L_{\perp}=5,10$ have a slightly bigger standard deviation, but as this discrepancy seems to be related to the small system thicknesses, we understand them as corrections to scaling, which are expected to be more prominent with $(+-)$ BC than periodic BC \cite{vasilyev_universal_2009}. This assumption can be further  tested by looking at the evolution of $L_{\perp}^{\beta/\nu} \rho^{-1} \sigma$ with $L_\perp$, Fig.~\ref{P_3D_BCMP}b). 
This reduced variance depends very little on $\rho$ but strongly on $L_\perp$. In the case of fully periodic boundary conditions, we interpreted the dependency on $\rho$ of the reduced variance as a signature of approaching the infinite slab limit. Here, the reduced variance depends very little on $\rho$ but strongly on $L_\perp$, thus the absence of collapse of the reduced probability distributions seems more likely to be related to corrections to the scaling limit rather than to the approach of the limit of slab geometry $\rho \to 0$.
The evolution of the reduced standard deviation is well captured by a fit of the form 
\beq
L_{\perp}^{\beta/\nu} \rho^{-1} \sigma = \hat{\sigma} + b_{eff} L_\perp^{-\omega_{eff}} \ ,
\label{power_law_fit_sigma}
\eeq
where $b_{eff},\o_{eff},\hat{\sigma}$ are fitting parameters for which we find 
$\omega_{eff} \approx 0.86$, $\hat{\sigma} \approx 0.35$ and $b_{eff}\approx 1.0$. 
This is consistent with a correction to scaling, as $\o_{eff}$ compares well with the exponent $\o=0.84(4)$  \cite{pelissetto_critical_2002} controlling the leading correction to scaling in the Ising model. A more thorough study of the impact of corrections to scaling on $\sigma$ would be required to confirm this point, but the analysis proposed here seems to  
convincingly  show that  $L_{\perp}^{\beta/\nu} \rho^{-1} \sigma$ approaches a constant value for large $L_{\perp}$, confirming the scaling proposed in Eqn. (\ref{sigmad}).

The kurtosis, our test of normality in the case of symmetric distributions, is always rather small and goes to zero as $\rho \to 0$ (Fig.\ref{P_3D_BCMP}c). Even for the smallest system $\rho=1$ and $L_\perp=5$, $\gamma_2$ is only $\approx 0.3$, which is a kurtosis we would expect in a system of approximately $\rho \sim 1/10$ with periodic boundary conditions. Thus, we obtain much more easily a Gaussian form with $(+-)$ BC. 
We stress here that there is no fundamental reason to expect a Gaussian form for finite $\rho$, as 
the physical origin of the unimodal distribution is different from the periodic boundary conditions case. The $(+-)$ BC forces the system to "demix" and the fluctuations of the order parameter come from the deformation of the interface between a $+$ oriented region and a $-$ oriented one. Hence, although the measured kurtosis appears small we cannot rule out a finite value for finite $\rho$.

The boundary conditions can be seen as local magnetic fields of value $\pm J$ acting on boundary spins. Lowering the value of this boundary field below a threshold one expects that the interface will disintegrate, giving  a crossover towards free boundary conditions (Dirichlet $(O,O)$). The distribution for free boundaries should be qualitatively similar to that for the fully periodic case, with a bimodal structure below $T_c$ \cite{binder_finite_1981} and so one should find a qualitative evolution in the distribution functions for finite $\rho$, as one passes this threshold. 

The limit case $L_\perp = 1$  here is a 2D system without boundary field, the two $+$ and $-$ boundaries canceling each other. As we are working at $T_c^{3D} > T_c^{2D}$, this 2D system in the limit $L_\perp = 1$ is at a temperature higher than the 2D critical one, giving rise to a unimodal distribution of the magnetization. This limit case ensures that with $(+-)$ boundary conditions we shall always observe a unimodal distribution, even when the limit $L_\perp \to 1$ is approached .

\subsection{Fixed $(++)$ boundary conditions}

%
\begin{figure*}[bht!]
\centering
{a) \hspace{-0.2cm}} 
\includegraphics[width=0.47\textwidth]{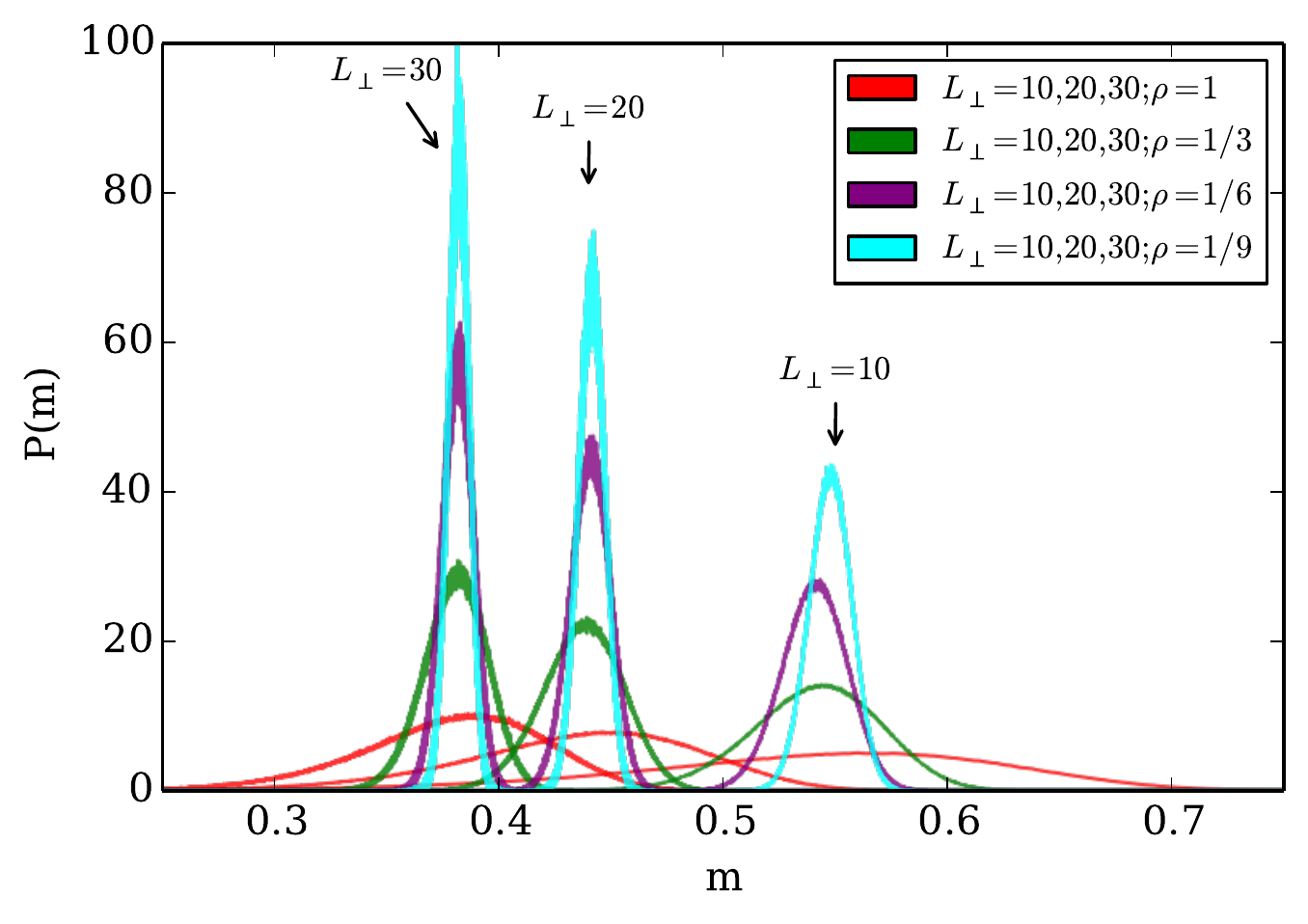} 
{b) \hspace{-0.2cm}} 
\includegraphics[width=0.47\textwidth]{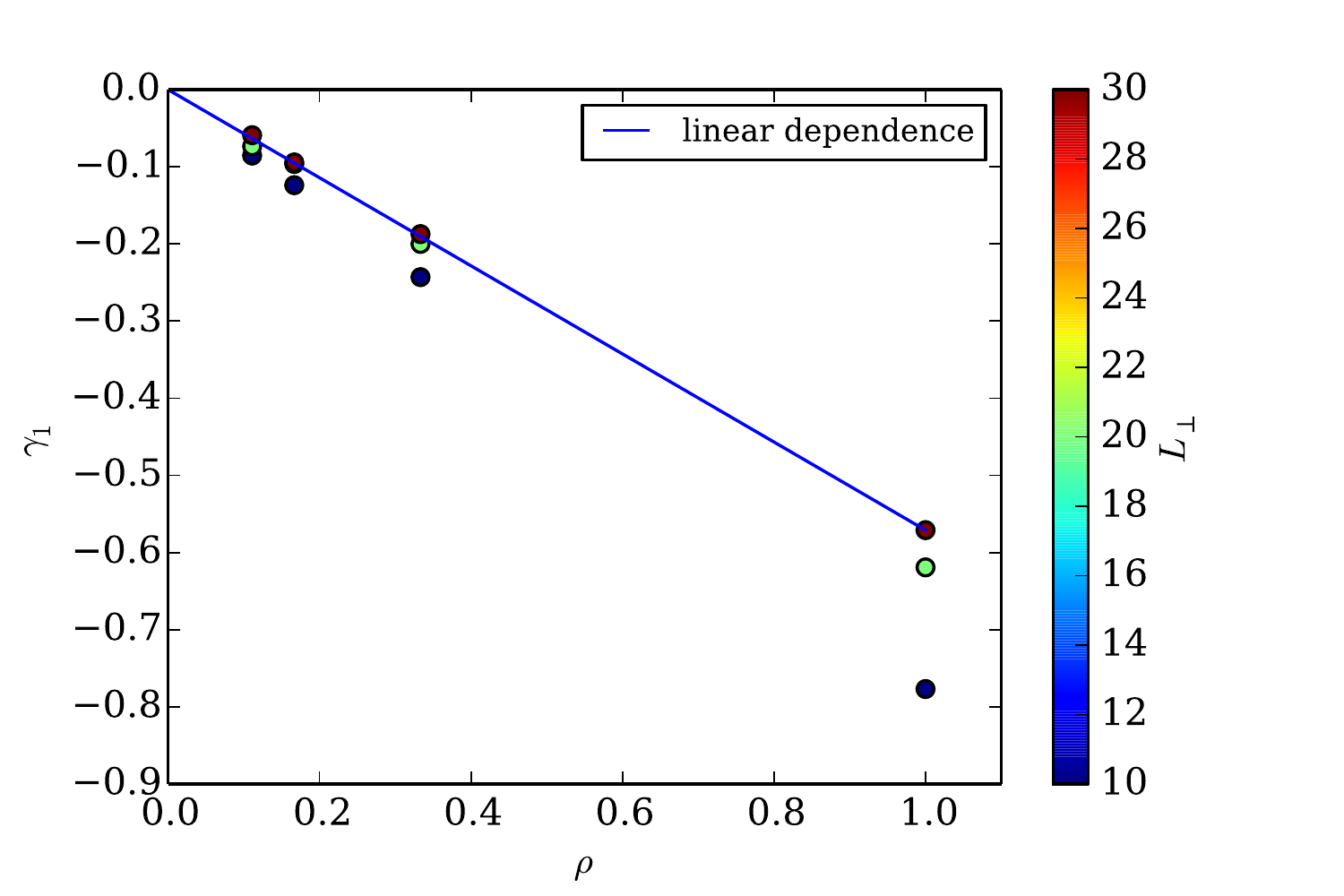}
\caption{ Results obtained by Monte-Carlo simulation of 3D Ising system with fixed $(++)$ BC for various system thicknesses $L_{\perp}$ and aspect ratios $\rho$ at the critical temperature $T=T_c^{3D}$.
{\bf a)}  Magnetization probability density $P(m)$ as a function of $m$. The distribution is always unimodal, but the symmetry breaking boundary conditions impose a non-zero average magnetization.
%
%
{\bf b)} Asymmetry of the distributions can be highlighted by looking at the skewness $\gamma_1$. We find the main trend to be $\gamma_1 \propto \rho$, represented by the continuous line, fitted on the data for the largest thickness $L_\perp=30$. Points color give $L_\perp$.
}
\label{BCPP_slab}
\end{figure*}
%
%
\begin{figure*}[!b!]
\centering
{a) \hspace{-0.2cm}} 
\includegraphics[width=0.47\textwidth]{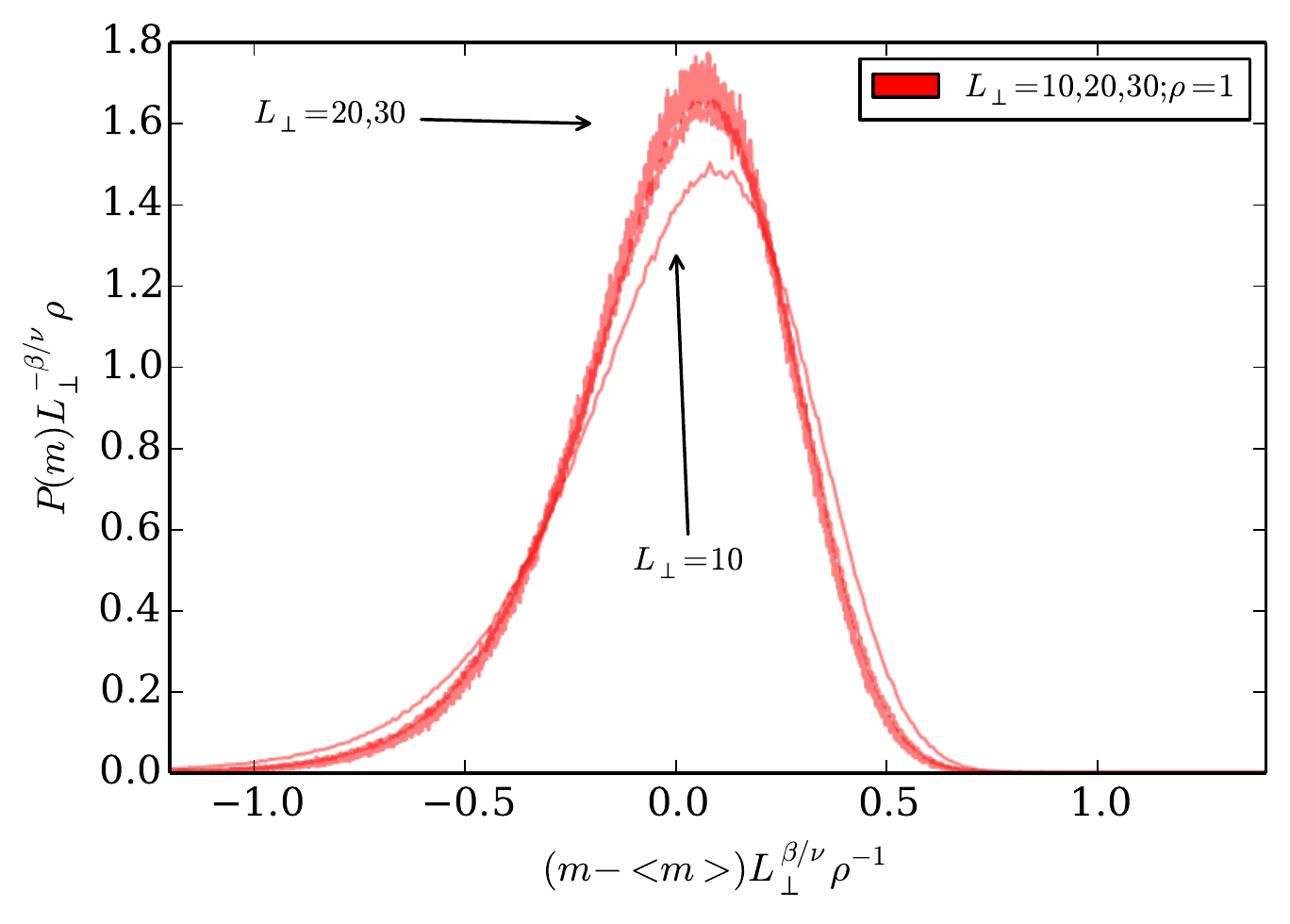}
{b) \hspace{-0.2cm}} 
\includegraphics[width=0.47\textwidth]{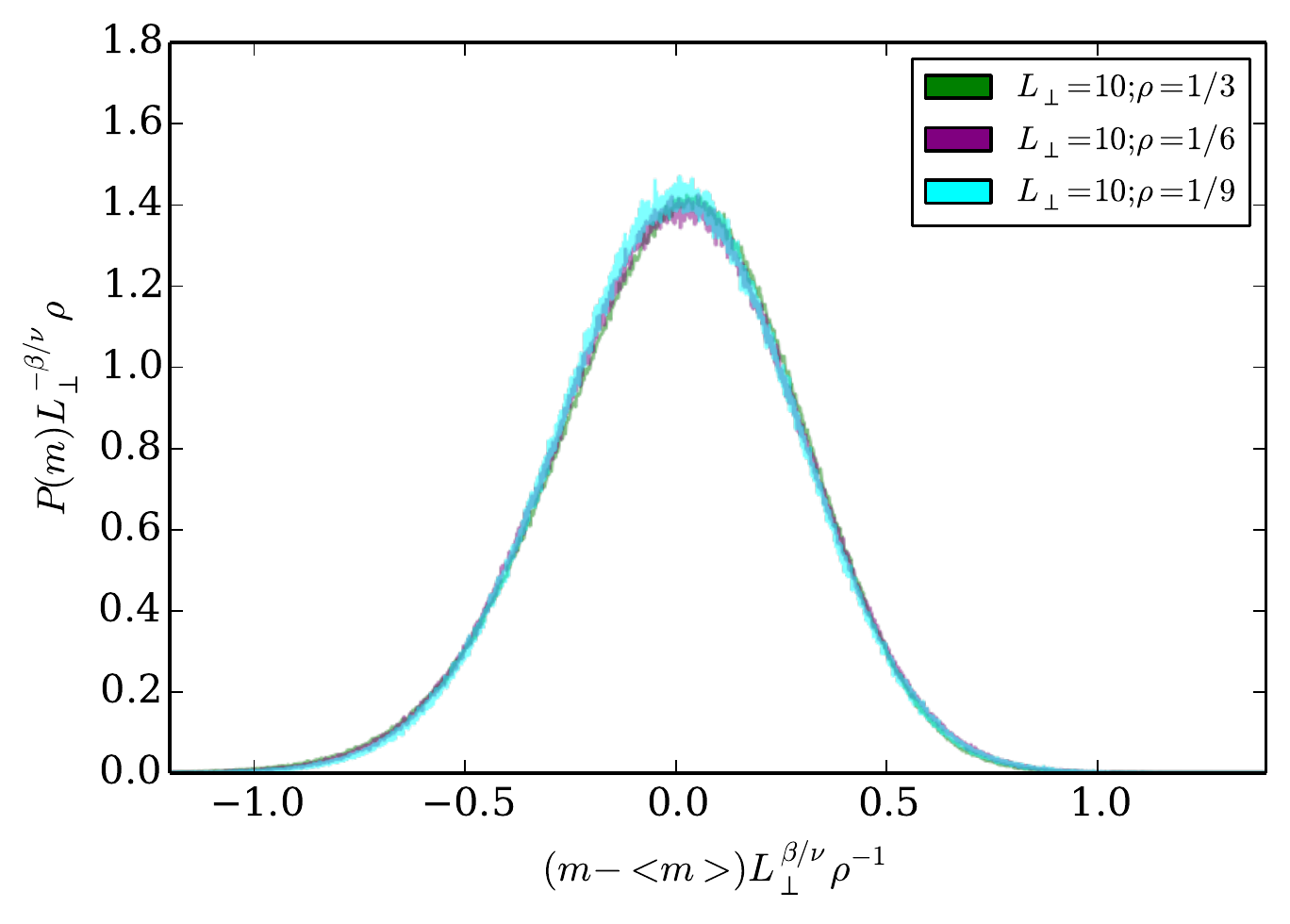} \\
{c) \hspace{-0.2cm}} 
\includegraphics[width=0.47\textwidth]{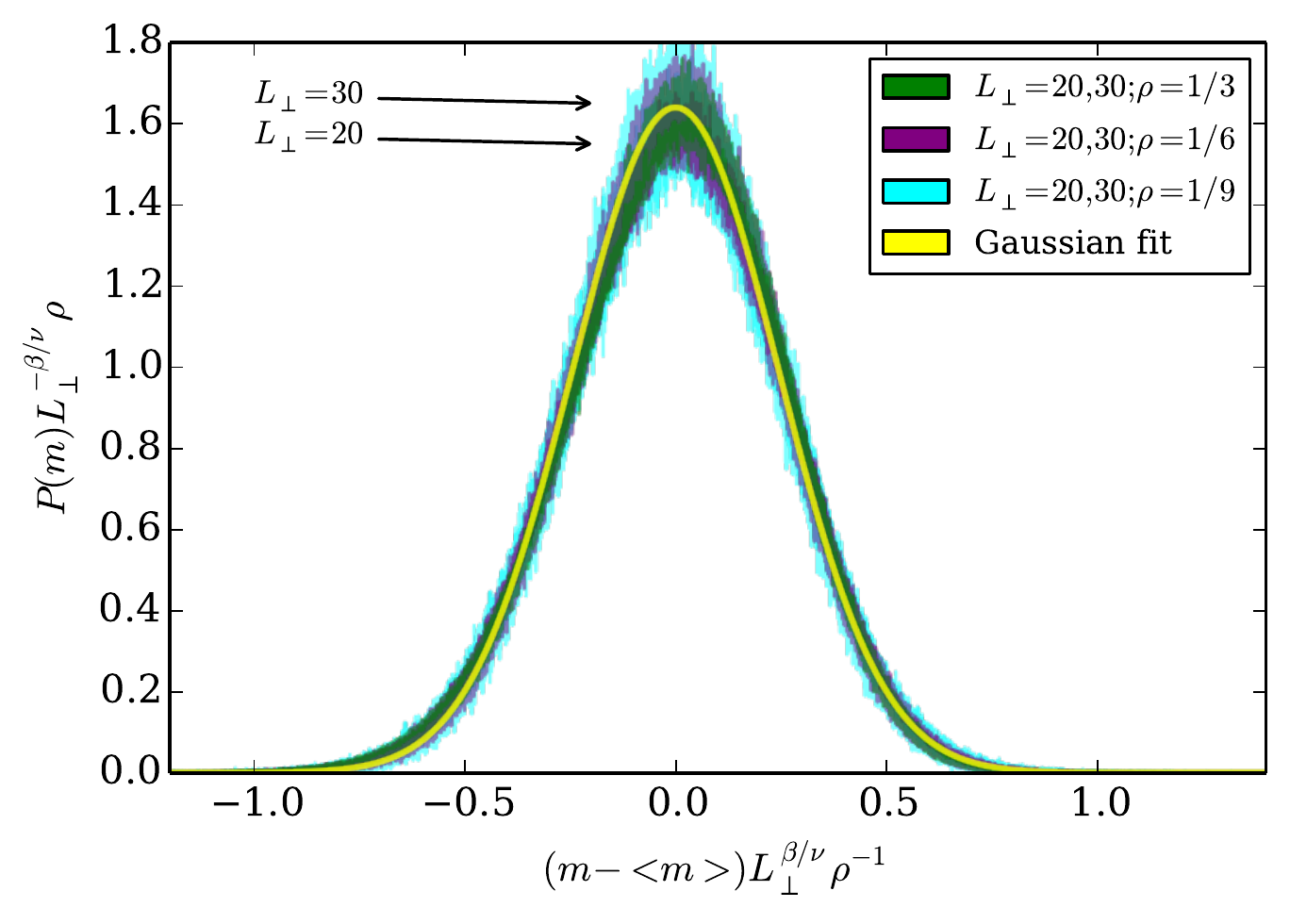}
{d) \hspace{-0.2cm}} 
\includegraphics[width=0.47\textwidth]{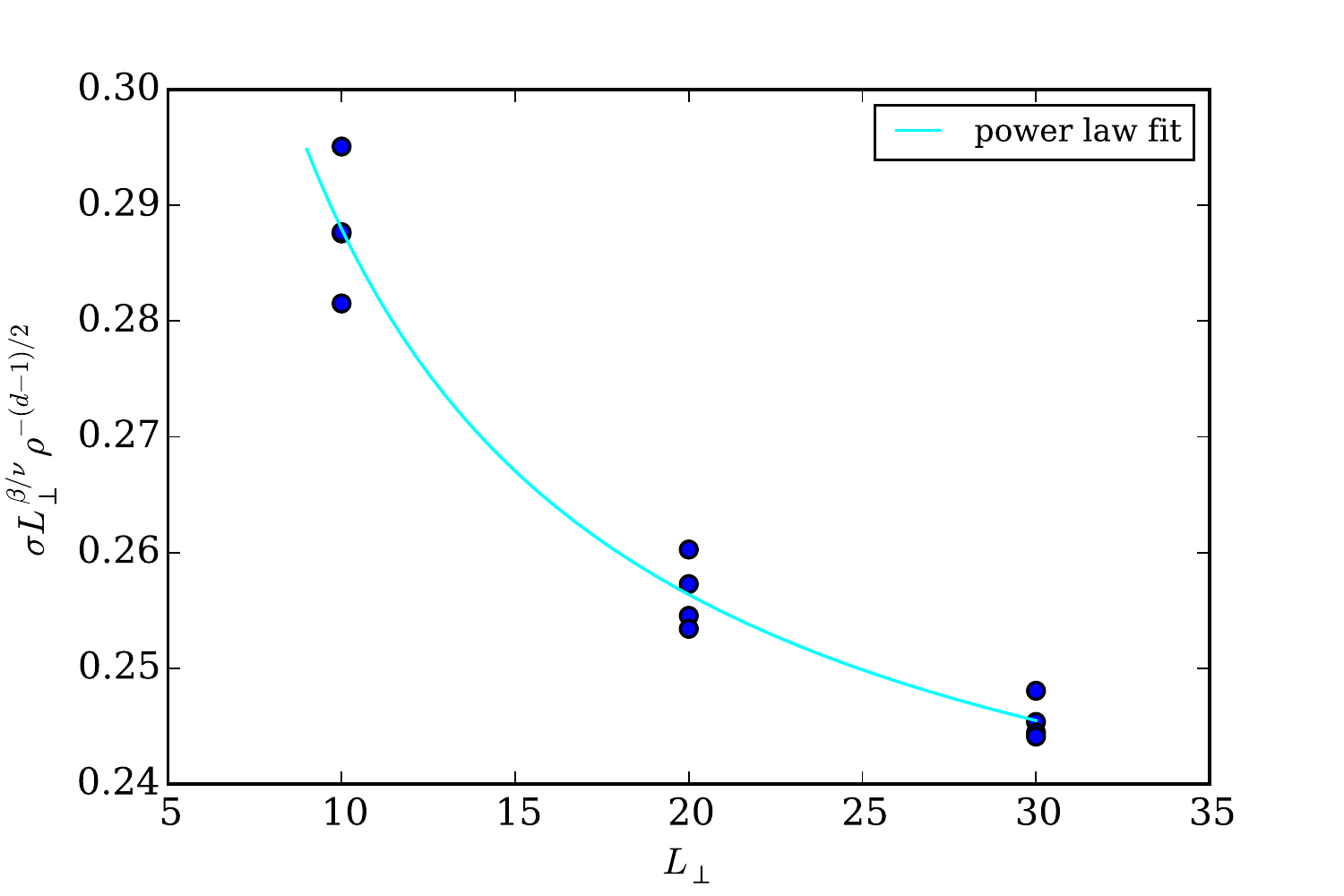}
\caption{ Results obtained by Monte-Carlo simulation of 3D Ising system with fixed $(++)$ BC for various system thicknesses $L_{\perp}$ and aspect ratios $\rho$ at the critical temperature $T=T_c^{3D}$.
{\bf a,b,c)} $L_{\perp}^{-\beta/\nu} \rho P(m-\langle m \rangle)$ versus $(m -\langle m \rangle)L_{\perp}^{\beta/\nu} \rho^{-1}$. For the sake of visibility, we have separated the most skewed distribution for $\rho=1$ {\bf (a)} and the smallest system thickness $L_{\perp}=10$ for which the variance is notably different from larger system thickness {\bf (b)}. For systems with $L_\perp \geq 20$ and $\rho <1$, we obtain a convincing collapse within the current precision {\bf (c)}. The continuous line is a Gaussian fit on data obtained with the largest system available.
{\bf d)} $\sigma L_\perp^{\beta/\nu}\rho^{-(d-1)/2}$ as a function of $L_\perp$: the scaled standard deviation in this case noticeably depends on both $\rho$ and $L_\perp$. The continuous line is a power law fit only intended as a guide to the eye.
}
\label{BCPP_slab_bis}
\end{figure*}
%
With symmetry breaking $(++)$ boundary conditions, the average magnetization $\langle m \rangle$ is no longer expected to be zero, the total imposed magnetic field being non-zero. As can be seen in figure \ref{BCPP_slab}a), the magnetization probability density is unimodal with the position $m_{max}$ of its maximum depending on the system size. The average magnetization $\langle m \rangle$ depends mainly on the thickness $L_{\perp}$ and only little on the aspect ratio, as expected given that the field is applied uniquely at the boundaries.

With the boundary conditions breaking the symmetry, the kurtosis is no longer the lowest order moment quantifying a difference from the Gaussian behavior. An asymmetry can be noticed in the distribution which can be quantified by  the skewness \cite{bramwell_magnetic_2001}:
\begin{equation}
\gamma_1=\langle \left( \frac{m- \langle m \rangle}{ \sigma} \right)^3 \rangle \ ,
\label{skewness}
\end{equation}
which strongly depends on the aspect ratio, Fig.~\ref{BCPP_slab}b). For a given thickness, the dependency of $\gamma_1$ with $\rho$ appears linear in the range investigated, as the fit to data for the largest system size studied, $L_\perp=30$, shows (see figure~\ref{BCPP_slab}b)). For finite $\rho$, $\gamma_1$ clearly remains non-zero in the scaling limit and appears to approach a value exceeding $0.5$ for the cubic system, although  $\gamma_1(L_{\perp})$ does not converge to its asymptote for the system sizes available. A more extensive study is required to extract  this asymptote.
As one approaches the slab limit, the system can be again divided into a large number of statistically independent sub volumes, ensuring that $\gamma_1 \to 0$ when $\rho \to 0$ as the central limit theorem applies. 

Centering the distribution by plotting $L_{\perp}^{-\beta/\nu} \rho P(m-\langle m \rangle)$ versus ${(m -\langle m \rangle)L_{\perp}^{\beta/\nu} \rho^{-1}}$, Fig.~\ref{BCPP_slab_bis}, allows us to compare fluctuations of the magnetization around the mean value for varying $\rho$ and $L_{\perp}$.
For clarity we have separated data for $\rho=1$ (Fig.~\ref{BCPP_slab_bis}a). In this case, the significant skewness, which is clearly visible, sets the distribution apart from those for other values of $\rho$, giving an aspect similar to that observed for the distribution of the amplitude of the magnetization in two and three dimensional XY systems \cite{bramwell_magnetic_2001}.
We also separate data for the smallest system thickness $L_{\perp}=10$ (Fig.~\ref{BCPP_slab_bis}b) for which the variance is notably different from larger system thickness. For systems with $L_\perp \geq 20$ and $\rho <1$ (Fig.~\ref{BCPP_slab_bis}c), 
the distributions approximate well to a Gaussian form, despite the finite value for the skewness. A Gaussian fit to the data
 for the largest system available gives $\hat{\sigma} = 0.25(1)$. The scaling form of Eq.\ref{scaling4} therefore seems to be approached as $L_\perp$ grows and $\rho$ decreases.
As for the case of fixed $(+-)$ boundaries, we test this further by examining the evolution of the scaled standard deviation $L_{\perp}^{\beta/\nu} \rho^{-1} \sigma$ with $L_\perp$, as shown in (Fig.~\ref{BCPP_slab_bis}d). Unlike the previous case we now find a strong dependence on both $\rho$ and $L_{\perp}$ making a quantitive test, including corrections to scaling difficult. However, it does appear that the scaled standard deviation will evolve to a constant value for larger $L_\perp$ and $\rho$ and it would certainly be interesting to examine this further in the future.

%

\section{Conclusion}

We have seen how both the aspect ratio and boundary conditions of an Ising system influence the fluctuations of the order parameter at the critical temperature. We used simple arguments to propose a scaling form in the limit of slab geometry $L_\parallel \gg L_\perp$, in which case the probability density tends to a Gaussian distribution in all studied cases. The data collapse was obtained by taking into account both the aspect ratio and the confinement length $L_\perp$.
Fully periodic, $(+-)$ and $(++)$ boundaries offer widely differing characteristics. In the case of fully periodic conditions the probability density function reflects the bulk $Z_2$ symmetry breaking of the Ising transition, evolving from a bimodal distribution for aspect ratio of unity, to gaussian in the limit of slab geometry. Imposing $(+-)$ boundary conditions ensures that the $Z_2$ symmetry  is topologically protected by the interface between $+$ and $-$ domains, resulting in a distribution function that is unimodal, even for an aspect ratio of unity and leading to a probability density function close to Gaussian for all values of $\rho$. Imposing $(++)$ boundaries breaks the symmetry, dictating the broken symmetry phase, ensuring  a non-zero value of the order parameter and giving a distribution function that depends both on the aspect ratio and the confining length scale, $L_{\perp}$.

With the ever increasing experimental control of nano-scale confinement, the understanding of how fluctuations develop in a confined system is of considerable theoretical and experimental importance. The $(+-)$ boundary conditions are experimentally relevant for wetting films \cite{binder_monte_2003,wilding_effect_1998} and for binary liquid mixtures stabilized in nano-scale wetting layers. These boundary conditions and $(++)$ boundaries are all relevant for magnetic thin films \cite{stanley_scaling_1999,vaz_magnetism_2008,jongh_experiments_2001}, where magnetic moments can be pinned at the surface. Here, the development of enhanced Gaussian fluctuations at the critical temperature are intimately related to the phenomenon of superparamagnetism, which could be used as a diagnostic for the variance of the probability density. The modification of the spectrum of critical fluctuations by confinement and boundary conditions gives rise to the critical Casimir force which has been the topic of an extensive literature in the past decades \cite{m._e._fisher_phenomenes_1978,gambassi_casimir_2009,lopes_cardozo_critical_2014}. The form of the Casimir force, like the fluctuation spectrum depends strongly on the boundary conditions and the two are closely related. In future work it would be interesting to extend the notions presented here to the study of magnetization profiles for different values of aspect ratios and confinement lengths, as well as to the depth dependence of local probability densities.


%

\medskip

\medskip

\noindent {\bf Acknowledgements}
We thank M.~Fruchart, C.~Charles, S. Ciliberto, F. Puosi, A. Rancon and T. Roscilde for useful discussions. The work was financed by the ERC grant OUTEFLUCOP and used the numerical resources of the PSMN at the ENS Lyon. P.C.W.H. acknowledges financial support from the Institut Universitaire de France.

\bibliographystyle{unsrt}
\bibliography{./biblio_zotero_these}

\end{document}